\newcommand{\BF}[1]{\mbox{\boldmath $#1$}}

\newcommand{\fBF}[1]{\mbox{\boldmath\footnotesize $#1$}}

\def\dddot#1{{\hspace{3pt}\dot{\phantom #1}\hspace{-2.4pt}\ddot{\!#1}}}

\documentstyle[eqsecnum,aps,epsf]{revtex}

\def\ins#1{}

\def\iems#1{}
\def\comment#1{}

\def\Det{{\rm  Det\,}}

\def\cm#1{}
\def\sbf#1{\mbox{\scriptsize{\bf #1}}}

\begin{document}
\setcounter{figure}{0}
\Roman{figure}

\title{{
Langevin Equation for Particle in Thermal Photon Bath
}}
\author{Z. Haba%
 \thanks{On leave from Institute of Theoretical Physics, University of
Wroclaw, Poland; e-mail: zhab@ift.uni.wroc.pl}
 and H. Kleinert%
 \thanks{Email: kleinert@physik.fu-berlin.de 
URL:
http://www.physik.fu-berlin.de/\~{}kleinert \hfill
} }
\address{Institut f\"ur Theoretische Physik,\\
Freie Universit\"at Berlin, Arnimallee 14,
14195 Berlin, Germany}
\maketitle
\begin{abstract}
The forward--backward path integral describing a
charged particle moving in a thermal bath of photons
is expressed in terms of the solution of a Langevin-type of equation.
Approximate methods for solving this equation are discussed.
\end{abstract}

%
\section{Introduction}
The interaction of charged nonrelativistic point
particles with radiation  is a fundamental problem
of atomic physics. The electrons in light atoms
 move with
nonrelativistic speed and may be  described by
 a Schr\"odinger equation  with the Hamilton operator
\begin{eqnarray}
    \hat H_{\sbf x} = \frac{1}{2M}\left[\hat{\bf p} - \frac{e}{c} {\bf A}({\bf x},t)\right]^2 +
  V ({\bf x}),
\label{Han}\end{eqnarray}
where ${\bf A} ({\bf x}, t)$ is the electromagnetic vector
potential and $ V({\bf x}) =Z  \alpha /r$ the Coulomb potential
of the nucleus. The decay of an atom is governed by dipole
radiation,
 and the matrix elements of the dipole operator
between initial and final atomic states allow us to calculate
directly the natural line width of a single atom in the vacuum.

In actual physical systems, this simple situation becomes
more involved.
An  atom may loose  its energy in a variety of competing
processes which lead to an additional broadening of its spectral
lines. In this note we want to set up a theoretical framework
for studying the broadening due to the interaction
of the atom  with a grand-canonical ensemble of
photons  in thermal equilibrium
at high and moderately high temperatures. This will be
referred to as a thermal {\em bath\/} of photons. Our work is a physically
more realistic version of the well-known
treatment of a particle in contact with
a thermal bath of oscillators \cite{FeynmanVernon,FeynmanHibbs,CaldeiraLeggett,PI}.
\section{Derivation of the Langevin equation}
The time evolution of the density matrix
$ \rho ({\bf x}_{+ a}, {\bf x}_{-a}; t)$ of the system
described by (\ref{Han}) follows
 a  quantum Liouville equation
\begin{eqnarray}
 i \hbar \partial_{t_b}  \rho  = \hat H_{{\sbf x}_{+a}}  \rho  - \hat H_{{\sbf x}_{-a}}
   \rho                 .
\label{Ham}\end{eqnarray}
This corresponds
to a global time evolution equation
\begin{eqnarray}
   \rho  ({\bf x}_{+a} , {\bf x}_{-a} ; t_b) = \int d {\bf x}_{+b}
 d{\bf x}_{-b}
   \,U \left( {\bf x}_{+b}   {\bf x}_{-b}  , t_b | {\bf x}_{+a}
    {\bf x}_{-a} , t_a\right)  \rho \left({\bf x}_{+b}
    , {\bf x}_{-b} , t_a\right).
\label{2.1}\end{eqnarray}
The evolution kernel
may be expressed as a forward--backward path integral
due to
Feynman and Vernon \cite{FeynmanVernon,FeynmanHibbs,CaldeiraLeggett,PI}:
\begin{eqnarray}
 U \left( {\bf x}_{+b}, {\bf x}_{-b}, t_b| {\bf x}_{+a} ,
   {\bf x}_{-a} , t_a  \right)
& = & \int   {\cal D}{{\bf x}_+}    {\cal D}{{\bf x}_-}
   \delta  \left({\bf x}_{+}   (t_b) - {\bf x}_{+b} \right)  \delta
      \left({\bf x}_{-}  (t_b) - {\bf x}_{-b}   \right) \nonumber \\
 &&\times   \exp \left\{  \frac{i}{\hbar}    \int_{t_a}^{t_b}
    \left[ \frac{M}{2} \left(\dot{\bf x}_+^2  - \dot{\bf x}_-^2 \right)
    - V ({\bf x}_+)  + V ({\bf x}_- )
    - \frac{e}{c} \dot{\bf x}_+  {\bf A} ({\bf x}_+)
     + \frac{e}{c} \dot{\bf x}_-  {\bf A} ({\bf x}_-)\right] \right\}.
\label{2.2}\end{eqnarray}
To describe a particle in a bath of photons,
we perform a thermal average
over the fluctuating vector potential $A ({\bf x},t)$.
This is done
with the help of
Wick's theorem, according to which the result can be expressed
completely in terms of the time-ordered
correlation functions at temperature $T$:
\begin{eqnarray}
 \langle\hat T \hat A^i ({\bf x},t) \hat A^j ({\bf x}', t')\rangle
\equiv G^{ij} ({ x}, { x}') & = &
   {\hbar c^2}
     \int \frac{d^3k}{2 \Omega _{\sbf k}(2\pi)^3}\delta ^{ij\,\rm tr}_{{\sbf k}{\sbf k}}
  \cos  {\bf k}
    ( {\bf x} -{\bf x}')
   \left[ \cos  \Omega _{\sbf k}  (t - t')
       \coth  \frac{\hbar \Omega _{\sbf k}}{2k_BT}
 - i \sin \Omega _{\sbf k}( t-t')\right]
\label{2.3}\end{eqnarray}
where    $  \Omega _{\sbf k}\equiv c|{\bf k}|$ are the frequencies
of wave vector ${\bf k}$,
  $k_B$  is the Boltzmann constant,
and
 $\delta ^{ij\,\rm tr}_{{\sbf k}{\sbf k}}\equiv
  \delta _{ij} -
   {k_i k_j}/{{\bf k}^2}$  the transverse $ \delta $-function,
We have used
the four-vector notation
$x \equiv ({\bf x},t) $  for space
and time. The operator $\hat T$ is the time-ordering operator.

At high temperature and for particle systems small in comparison
to the mean wave length $1/|{\bf k}|$, we may neglect the ${\bf x}$-dependence
in $G(x,x')$.
Then $G(x,x')$ can be approximated  by
\begin{equation}
 G(x,x')\approx \frac{2}{3}\frac{\hbar }{4 \pi c}  \left\{\frac{k_BT}{\hbar } \left[\delta (t-t')-
\frac{\hbar^2}{12(k_BT)^2} \ddot\delta (t-t')-\dots\right]-
i \dot  \delta (t-t')\right\},
\label{@Gexp}\end{equation}
%
the brackets being the Fourier transform of
the Taylor expansion of  $({\hbar \Omega _{\sbf k}}/{2k_BT})
 \coth  {\hbar \Omega _{\sbf k}}/{2k_BT}
$ in (\ref{2.3}).
Moreover, the fluctuations
 around the average path
${\bf x}(t) \equiv [{\bf x}_+ (t) + {\bf x}_- (t)]/{2}$
are small, so that we can approximate
\begin{eqnarray}
 V({\bf x}_+)  - V({\bf x}_-)   \simeq
{\bf y\cdot}  \BF\nabla V ( {\bf x})
,
\label{}\end{eqnarray}
where ${\bf y}(t)\equiv
 {\bf x}_+(t) - {\bf x}_- (t)$ is the difference between forward and backward paths.
Then the path integral (\ref{2.2})
 takes the form (see \cite{HabaKleinert} for more details)
\begin{eqnarray}
 U ({\bf x}_{+b} , {\bf x}_{-b}, t_b |
      {\bf x}_{+a} , {\bf x}_{-a}, t_a)
   &=& \int {\cal D }{\bf x}     {\cal D }{\bf y}
        \delta  \left({\bf x} (t_b) - {\bf x}_b
	 \right)  \delta  \left({\bf y} (t_b) - {\bf y}_b
   \right)
   \nonumber \\
&\times	   &	       \exp \left\{  \frac{i}{\hbar} \int^{t_b}_{t_a}dt
 \left[ \dot{\bf y} \left( M\dot{\bf x} -  M\gamma \ddot{\bf x}
   + i \frac{w}{2 \hbar  } \dot{\bf y}
-i\frac{w\hbar }{24 (k_BT)^2}\dddot{{\bf y}}+\dots \right)
 - {\bf y} {\BF \nabla} V({\bf x})\right] \right\}  ,
\label{2.4}\end{eqnarray}
where we have abbreviated
\begin{eqnarray}
    \gamma \equiv  \frac{e^2}{3 \pi c^3M} , ~~~
  w\equiv 2M  \gamma k_BT
\label{@}\end{eqnarray}
At $t_a$, the paths ${\bf x}(t)$
and
${\bf y}(t)$
 start from ${\bf x}_a\equiv
({\bf x}_{+a}+{\bf x}_{-a})/2$
and
${\bf y}_a\equiv {\bf x}_{+a}-{\bf x}_{-a}$,
respectively.
Representing the $ \delta $-function $\delta  \left({\bf y} (t_b) - {\bf y}_b
   \right)               $    in (\ref{2.4})
as a Fourier integral, and inserting
for ${\bf y}(t_b)$ the
equation
\begin{eqnarray}
{\bf y}(t_b) = \int^{t_b}_{t_a}dt'\, \dot{\bf y}(t' )  + {\bf y}_{a},
\label{2.5}\end{eqnarray}
we rewrite the evolution kernel
in the form
\begin{eqnarray}
\lefteqn{\!\!\!\!\!\!\!\!\!\!\!\!\!
U \left( {\bf x}_{+b}, {\bf x}_{-b}, t_b | {\bf x}_{+a}, {\bf x}_{-a}, t_a \right)
 }\nonumber \\
 &&= \int \frac{d^3p }{(2 \pi )^3} \int {\cal D } {\bf x}
{\cal D } {\bf y}   \exp
  \left\{  \frac{i}{\hbar} \int^{t_b}_{t_a} dt \left[\dot{\bf y}
     \left({\BF  \eta } + \frac{iw}{2\hbar} \dot{\bf y} +\dots
      \right) - {\bf y}_b  {\BF \nabla}
  V({\bf x})\right]  - \frac{i{\bf p} }{\hbar}
  ({\bf y}_b -{\bf y}_a) \right\}
  \delta  \left({\bf x}(t_b) - {\bf x}_{b}\right) ,
\label{2.6}\end{eqnarray}
where we introduced the new variable
\begin{eqnarray}
{ \BF\eta}  (t) \equiv  M \dot{\bf x} (t) - M \gamma \,
\ddot{\bf x}  (t) + \int^{t}_{t_a} dt' \BF\nabla
 V \left({\bf x}(t')\right)
-{\bf p} .
\label{2.7}\end{eqnarray}
In Eq.~(\ref{2.6}),
the
variables
$\dot{\bf y}(t)$
at different  points $t_a < t < t_b$
are independent  of each other,
and we
 choose at the end points $\dot{{\bf y}} (t_a) = \dot{\bf y }
 (t_b) = 0$. This may be justified in
 a time sliced formulation, where the integrations over the variables
next to the end points
give only a trivial factor with respect
to the product of integrals
in which these variables are held fixed
at the endpoint values.
Furthermore,
the path integral (\ref{2.6}) does not depend
on $\BF\eta(t)$ outside the interval $t\in(t_a,t_b)$, and is independent of
$\BF \eta (t_b)$ and
$\BF \eta (t_a)$.
Hence we may choose
$ {\BF \eta} (t_a) =
 {\BF \eta} (t_b) =0$, for convenience,
and Eq.~(\ref{2.7}) can be solved
as a differential equation on the interval
$t_a \leq t \leq t_b$
\begin{eqnarray}
 M \ddot{\bf x}  - M \gamma\!\dddot{{\bf x}}  + \BF\nabla V ({\bf x})
= \dot{\BF \eta} (t) ,
\label{2.8}\end{eqnarray}
with the initial conditions
\begin{eqnarray}
&& {\bf x} (t_a) = {\bf x}_{a},
~~~~
M \dot{\bf x} (t_a) - M \gamma  \ddot{\bf x} (t_a) = {\bf p} .
\label{2.8a}\end{eqnarray}
We now  perform the integral over $\dot{\bf y}$
 in Eq.~(\ref{2.6}). We shall, from now on,
neglect the
expansion terms indicated by the dots,  and obtain the path integral
\begin{eqnarray}
\lefteqn{\!\!\!\!\!\!\! U \left( {\bf x}_{+b}, {\bf x}_{-b}, t_b | {\bf x}_{+a},
 {\bf x}_{-a}, t_a \right)  }\nonumber \\
 &&= \int \frac{d^3p }{(2 \pi )^3} \int {\cal D } {\bf x}
\exp \left[ - \frac{1}{2w}
   \int^{t_b}_{t_a}dt\, {\BF \eta}^2 (t) \right]
   \exp
  \left\{ - \frac{i}{\hbar}\left[  {\bf y}_b  \int^{t_b}_{t_a} dt\,
  {\BF \nabla}
  V(x) + {\bf p} ({\bf y}_b -{\bf y}_a)\right]\right\}
  \delta  \left({\bf x}(t_b) - {\bf x}_{b}\right)
\label{2.9}\end{eqnarray}
where ${\BF \eta}(t)$ depends on ${\bf x}(t)$ via (\ref{2.7}).
There are some virtues of this representation in comparison
with the path integral (\ref{2.6}), in particular,
if forward and backward paths start out and end at the same
 points,  such that ${\bf y}_a={\bf y}_b={\bf 0}$: the
oscillatory integral in (\ref{2.6})
is transformed into a Gaussian integral
(\ref{2.9}) which converges exponentially fast.
Such  a representation is obviously
more suitable for numerical simulations.

For convenience, let us express the evolution of the density matrix (\ref{2.1})
in terms of the Wigner function defined by the Fourier transform
\begin{eqnarray}
 W ({\bf x}, {\bf p};t) = \left(\frac{1}{2 \pi \hbar }\right)^{3}
  \int d^3y\, e^{i {\sbf p} {\sbf y}/\hbar}
 \rho \left({\bf x}, {\bf y};t\right).
\label{2.10}\end{eqnarray}
Here and in the sequel, we omit subscripts $b$
from $t$ for brevity.
Then Eq.~(\ref{2.9}) yields the time
evolution equation for the Wigner function as
a functional integral
%
\begin{eqnarray}
W ({\bf x}, {\bf p}; t) &  =  &
 \int {\cal D } {\bf x} \exp \left[ - \frac{1}{2w}
 \int^{t}_{t_a} dt' \, {\BF \eta}^2(t') \right]
 W \left({\bf x} (t), {\bf p} -   \int^{t}_{t_a} dt'\,
     \BF\nabla V({\bf x}(t' )); t_a\right).
\label{2.11}\end{eqnarray}
This equation
 has a simple physical interpretation.
In the limit $T \rightarrow    0$, the functional integral
(\ref{2.11}) is concentrated around $\BF\eta(t')\equiv {\bf 0}$,
corresponding to  a deterministic solution
of Eq.~(\ref{2.7}) with $\BF\eta(t')\equiv {\bf 0}$.
In this limit, we obtain
from Eq.~(\ref{2.11})
 the Wigner function
\begin{eqnarray}
 W ({\bf x}, {\bf p};  t) = W \left({\bf x} (t), {\bf p} -
  \int^{t}_{t_a}  dt' \,\BF\nabla V({\bf x} (t')); t_a\right).
\label{WF}\end{eqnarray}
 If the particle is decoupled from the bath,
 $\gamma =0$, we have    ${\bf p} = M \dot{\bf x} (t_a)$
and
$ M \dot {\bf x}(t)= {\bf p} - \int^{t}_{t_a} dt\, \BF\nabla V ({\bf x} ( t)) $,
and we see that
the time evolution of the Wigner function
is given by
the Liouville equation
\begin{eqnarray}
 \partial_{t} W + \frac{1}{M}{\bf p}\cdot \BF\nabla_{{\sbf x}} W
 - \BF\nabla_{\sbf x}
V\cdot {\BF\nabla}_{\sbf p}
 W = 0      .
  \label{@}\end{eqnarray}

The
time evolution kernel
can also be expressed as
a path integral
over the noise variable ${\BF  \eta }(t)$.
For this simply change the integration  variable
in (\ref{2.11}) from ${\bf x}(t)$ to ${\BF  \eta }(t)$.
To find the functional determinant, we
integrate Eq.~(\ref{2.7}) once more and write
the result as
\begin{eqnarray}
 \partial_t^{-2} \dot{\BF \eta} (t) &= &
 M[ {\bf x}(t) -  {\bf x}_{a}]
    - M  \gamma [\dot{\bf x}  (t) - \dot{\bf x}
      (t_a)] + \int^{t}_{t_a} ds \int_{t_a}^{s} ds'\,
 {\BF \nabla }V({\bf x} (s'))
-{\bf p}   (t-t_a) .
\label{2.12}\end{eqnarray}
Differentiating this we obtain
\begin{eqnarray}
    \frac{ \delta \dot\eta_i(t)}{ \delta x_j(t')} =    M\partial_t^2 \left[
 \delta _{ij}
  \delta  (t-t') -  \gamma   \delta _{ij}
\dot\delta (t-t')
    +\frac{1}{M} \int^{t}_{t_a}  ds \int_{t_a}^{s} ds'\,
	      { \nabla } _i
	      { \nabla } _j
 V ({\bf x}(s'))  \delta (s' - t')\right].
   \label{2.13}\end{eqnarray}
 Applying the product formula ${\Det AB } = \Det A\, \Det B$,
the identity
$
  \Det (1+K) = \exp \left({\rm Tr}\, K - \frac{1}{2} {\rm Tr}\, K^2 \dots \right),
$
and the property
$ {\rm Tr}\, K^n=0$,
we see that the Jacobian
for the transformation
$ {\bf x}(t)\rightarrow  {\BF \eta}(t)  $ is a constant, so that
  we can rewrite Eq.~(\ref{2.11})
 as
\begin{eqnarray}
 W \left({\bf x}, {\bf p}; t\right) = \left\langle W
	\left({\bf x} (t) , {\bf p} - \int^{t}_{t_a}  dt'\,
 {\BF \nabla } V ({\bf x}(t' )), t_a \right)\right\rangle_{{\fBF  \eta }},
\label{@www}\end{eqnarray}
where the average
with respect to ${\bf  \eta }(t)$ fluctuations is performed
with the functional integral
\begin{eqnarray}
 \left\langle ~\dots~
\right\rangle_{{\fBF  \eta }}\equiv
 \int {\cal D } {\bf x}~\dots~ \exp \left[ - \frac{1}{2w}
 \int^{t_b}_{t_a} dt \, {\BF \eta}^2(t) \right] .
\label{@}\end{eqnarray}
The calculation of (\ref{@www}) proceeds by
solving first the Langevin equation
(\ref{2.8}) with the boundary conditions (\ref{2.8a})
to obtain the solution ${\bf x}(t)$, and subsequently take the
expectation value with respect to the white noise
with the correlation function
\begin{eqnarray}
 \langle \eta^i (t) \eta^j (t') \rangle =  w\,  \delta ^{ij} \delta
 (t-t') .
\label{2.15}\end{eqnarray}
The
quantum corrections to the Langevin equation
are taken into account
by
 replacing
(\ref{2.15}) by  the
colored-noise correlation function
\begin{eqnarray}
\langle \eta^i (t) \eta^j (t') \rangle & = & {w}  \delta ^{ij}
{\rm coth} \left(\frac{i\hbar}{2k_B T} \frac{d}{dt}\right)
  \frac{i\hbar}{2k_BT} \frac{d}{dt}  \delta (t-t') \nonumber\\
 & = & w  \delta ^{ij} \left[ 1 - \frac{\hbar^2}{12 (k_BT)^2
} \frac{d^2}{dt^2}- \dots \right]  \delta (t-t')
\end{eqnarray}

\section{Approximations to Liouville equation}
                    From
 Eq.~(\ref{2.15})
we see  that the average size of $\BF\eta$ is  $ \sqrt{2M \gamma k_B T}$ and thus proportional to  $\sqrt{T} $.
Let us therefore set
$
  \eta =  \sqrt{T/T_ H}  \tilde\eta$
with the characteristic Bohr temperature
$T_ H \equiv  \alpha ^2Mc^2/ k_B~( \alpha \equiv e^2/\hbar c)$.
Restricting ourselves now to one-dimensional
systems, we search  a solution of Eq.~(\ref{2.8})
 in the form
\begin{eqnarray}
 x(t) = e^{-  \gamma B(t)}\left[
x_{\rm cl} (t) +  \sqrt{T/T_H} Q(t)\right] ,
\label{3.2}\end{eqnarray}
where the first term on the right-hand side
describes small dissipative correction to the classical equation
\begin{eqnarray}
 M \ddot x_{\rm cl} + V' (x_{\rm cl}) = 0.
\label{3.3}\end{eqnarray}
The second term
on the right-hand side
depends on the noise $\tilde \eta$.
Then, comparing terms of first order in $ \gamma $
or $\sqrt{T/T_ H}$, and considering  terms $ \gamma \sqrt{T/T_ H}$
 as being of higher
order, we obtain the following equations
\begin{eqnarray}
 x_{\rm cl} \ddot B + 2 \dot B \dot x_{\rm cl} + B \left[\ddot x_{\rm cl}
 + \frac{1}{M} V'' (x_{\rm cl})
x_{\rm cl}\right] -\frac{1}{M}
  V'' (x_{\rm cl}) \dot x_{\rm cl} = 0  ,
\label{3.4}\end{eqnarray}
\begin{eqnarray}
 M \ddot Q + V'' (x_{\rm cl}) Q = \dot{\tilde{\eta}} .
\label{3.5}\end{eqnarray}
The solution $x_{\rm cl}(t)$ depends on the boundary condition
(\ref{2.8a}), in which we set $ \gamma =0$ in the present
lowest-order approximation.

The first equation is closely related to the second
as can be seen by
 introducing  $A(t)\equiv x_{\rm cl}(t)B(t)$ which satisfies
\begin{equation}\label{3.5a}
M\ddot A+V^{\prime\prime}(x_{\rm cl})A=
V^{\prime\prime}(x_{\rm cl})\dot x_{\rm cl}.
\end{equation}
Let $G(t,t')$ be the Green function of the operator
on the left-hand sides of (\ref{3.5}) and (\ref{3.5a}), satisfying
the harmonic differential equation
\begin{equation}
\left[ \frac{d^{2}}{dt^{2}}+\Omega ^2(t)\right]
G(t,t')= \delta (t-t'),
\label{@geq}\end{equation}
with the time-dependent frequency
\begin{equation}
 \Omega ^2(t)\equiv
\frac{1}{M}V^{\prime\prime}(x_{\rm cl}(t)).
\label{@}\end{equation}
Then, the solutions of Eqs.~(\ref{3.4}) and (\ref{3.5})
are given by
\begin{equation}
Q(t)=\int_{0}^{t}dt'\, G(t,t')\,
\dot{\tilde{\eta}}(t')
\end{equation}
and
\begin{equation}
B(t)=\frac{1}{Mx_{\rm cl}(t)}\int_{0}^{t}dt'\, G(t,t') \,
V^{\prime\prime}(x_{\rm cl}(t')) \dot x_{\rm cl}(t'),
\end{equation}
respectively.
The Green function can be expressed in terms of two independent
solutions $\xi_1(t)$ and $\xi_2(t)$
of the
homogeneous version of the
differential equation
(\ref{@geq}) \cite{klch,klchhab,klchhbib}
 with the initial conditions
$\xi_1(0)=0$ and $\dot \xi_1(0)=1$
and $\xi_2(0)=1$ and $\dot \xi_1(0)=0$.
These, in turn, can be obtained directly
from the classical solution $x_{\rm cl}(t)$\cite{4.99}.

 If the classical solutions are bounded in all directions
 of phase space, then the Green function
 and its derivative are bounded. In such a case it follows
 from Eq.~(3) in Ref.~\cite{klchhab} that the random perturbation grows at most
 linearly in $tT/T_H$.
A  deviation of magnitude $D$ from the classical
 solution appears at the time which is proportional to
 $ D/T$.
  However, if the classical system
 is chaotic, which
 can happen either  in multidimensional systems or in
 a one-dimensional system with a time-dependent potential,
 then
the classical solutions
 as well as the Green function can grow exponentially fast
 in some directions of the phase space.
In these directions,
 the classical solution expressed in terms of
 $\sqrt{T/T_ H}Q$ can grow to a
size of the order
 $D$ in a much shorter time $\ln(D/T)$.
 Such
an increase of quantum corrections
in chaotic systems
 has recently been discussed by H.Zurek\cite{Zu}.
For the
quantum  systems in a photon bath under discussion,
 this result  is modified by the
 damping factor $e^{-B(t)}$. Now, $B(t) $ can be bounded, or it can increase
 linearly in $t$, as it happens for
a linear oscillator.
In a chaotic system,
 $B(t)$ can also grow exponentially fast .
Such a strong
 friction dampens completely
the chaotic growth of quantum corrections, such that there
is no need to
 apply quantum mechanics to large macroscopic systems. It behaves
 like a strongly damped classical system.

The solution
(\ref{3.2})
 may be inserted into the
Eq.~(\ref{2.11})
to find the time evolution
of the Wigner function.

Our semiclassical
expansion  can  be extended systematically to any
order, albeit with increasing complexity.
Perturbative as well as full nonperturbative solutions can
be found most efficiently on a computer,
yielding the full time
evolution of the
 Wigner function and thus of the density matrix.
\section{Harmonic Potential}
The solution of our euqtions is simple
for a
harmonic potential
\begin{eqnarray}
 V (x)=  \frac{M \omega ^2}{2} x^2  .
\label{4.1}\end{eqnarray}
 Then
Eqs.~(\ref{3.4}) and
Eq.~(\ref{3.5})
have the solutions
\begin{eqnarray}
  B(t) = \frac{ \omega ^2}{2} (t-t_a),~~~~
 Q(t) = \frac{1}{M \omega } \int^{t}_{t_a} dt'\,\sin  \omega (t-t' )\,\dot{\tilde\eta} (t' ) ,
\label{4.3}\end{eqnarray}
  and (\ref{3.2}) yields to lowest order in  $\gamma $
the orbit
\begin{eqnarray}
 x(s) & = &e^{ - \gamma  \omega ^2 (s-t_a)/2}
  \left\{ x_{a} \cos \omega (s-t_a)
+ \frac{p}{1 +  \gamma  \omega }
   \frac{1}{M\omega}
\,\sin \omega (s-t' )
+   \frac{1}{M\omega}
\int^{s}_{t_a} dt'\,\sin \omega (s-t' )
 \dot \eta(t' ) \right\}
\label{4.4}\end{eqnarray}
which determines the Wigner function via Eq.~(\ref{WF}).

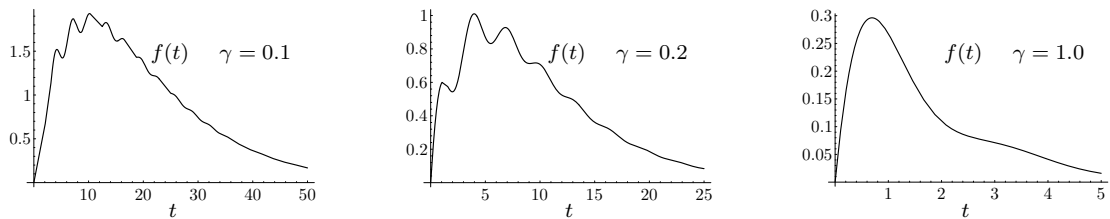
\begin{figure}[tbhp]
\input pp10.tps  ~\\[1mm]
\caption[]{Time dependence of
fluctuation width in Eq.~(\ref{4.6}) for different values of friction constant $ \gamma $.
Time is measured in units of $1/ \omega $.
 }
\label{pp10}\end{figure}

The fluctuation width is
\begin{eqnarray}
 \Big\langle \big( x(t) - \langle x(t)\rangle\big)^2\Big\rangle =
 \frac{w}{2M^2 \omega } f_ \gamma ( \omega t)\equiv
 \frac{w}{2M^2} (t-t_a) e^{-
 \gamma  \omega ^2(t-t_a)}
 \left[ 1 + \frac{\sin 2  \omega  (t-t_a)}{2 \omega (t-t_a)}
\right] .
\label{4.6}\end{eqnarray}
  For small times, this shows the same
linear
growth in time
 as
for the Brownian motion of a free particle with $ \omega =0$:
\begin{eqnarray}
 \Big\langle \big( x(t) - \langle x(t)\rangle\big)^2\Big\rangle =
 \frac{w}{M^2}  (t-t_a) ,
\label{4.5}\end{eqnarray}
As the time grows, the width oscillates around this
behavior with frequency $ 2\omega$.
For large times of the order $ 1/ \gamma \omega ^2$,
on the other hand,
the width goes
exponentially fast to zero (see Fig.~\ref{pp10}).

In the free-particle limit,
the solution (\ref{3.2})
reduces to
\begin{equation}
x(t)=x+\frac{p}{M}(t-t_a)+\frac{1}{M}\int _{t_a}^t dt'\, \eta (t').
\label{4.0}\end{equation}
Inserting this
into formula (\ref{2.11}), we obtain
the evolution
of the Wigner function.
For a wave packet with momentum $k$
and position $x$,
the result is
\begin{eqnarray}
 W(x, p\,; t_a) = \frac{1}{\pi\hbar} \exp \left[- \frac{(p-k)^2
\sigma }{\hbar^2}
  - \frac{(x-\bar x)^2}{ \sigma } \right].
\label{4.7}\end{eqnarray}
Inserting (\ref{4.0}) into (\ref{2.11}),
we obtain for $W$ the formula (\ref{4.7}) with the replacements
\begin{equation}
\bar x \rightarrow\bar x(t)= \bar x - \frac{p}{M} (t-t_a)
~~\mbox{and} ~~  \sigma \rightarrow  \sigma (t)=\sigma \left[  1 + \frac{2w}{M^2}
 (t-t_a)\right],
\label{@}\end{equation}
i.e.,~a free evolution with the well-known spreading of the wave packet.

For  the oscillator potential,
the  mean position and the mean momentum
of the particle run along the corresponding classical trajectories,
and the thermal spread behaves as in
(\ref{4.6}).

The spread in momentum space can be calculated from this using the
 relation
\begin{equation}
 \langle \left( p(t) - \langle p(t)\rangle\right)^2 =
M^2  \omega ^4 \left\langle \int^{t}_{t_a} \left(x(s) -
 \langle x(s)\rangle \right)^2  \right\rangle ds.
\label{@}\end{equation}
\\~\\
This research is supported by a grant from a governmental
 German university
 program
HSP III.


\begin{thebibliography}{11}
%
\bibitem{FeynmanVernon}
R.P.~Feynman and F.L.~Vernon, Ann.~Phys.~(N.Y.) {\bf 24}, 118 (1963)
\bibitem{FeynmanHibbs}
R.P.~Feynman, A.R.~Hibbs, {\em Quantum Mechanics and Path
Integrals\/}, McGraw Hill, New York 1965
%
 \bibitem{CaldeiraLeggett}
 A.O.~Caldeira and A.J.~Leggett, Ann.~Phys.~{\bf 149}, 374
(1983), {\bf 153}, 445(E) (1984).
%
\bibitem{PI}
H.~Kleinert,
{\em Path Integrals and Quantum Mechanics, Statistics, and
Polymer Physics\/}, 2nd edition, World Scientific, Singapore, 1995.
%
%
\bibitem{HabaKleinert}
 Z.~Haba, H. Kleinert,
{\em Master equation for electromagnet dissipation
and decoherence of density matrix\/}, Berlin preprint 2000
%
\bibitem{klch}
{H. Kleinert} and {A. Chervyakov},
Phys. Lett. A {\bf 245\/}, 345 (1998) (quant-ph/9803016);
 J. Math.  Phys. B {\bf 40\/}, 6044 (1999) (physics/9712048).\\
For more details see Sections 3.3.1, 3.5,  3.21, and 4.3 in the third edition
of Ref.~\cite{PI}
which can be downloaded
from
http://www.physik.fu-berlin.de/\~{}kleinert/b3.
\bibitem{klchhab}
 Z. Haba, Lett. Math. Phys. {\bf 47}, 321 (1999).

\bibitem{4.99}
See Eq.~(4.99) in
the third edition
of Ref.~\cite{PI}.


\bibitem{klchhbib}
L. Bieberbach, {\em Einfhrung in die Theorie
der Differentialgleichungen in Reellen Gebieten\/},
Springer, 1956.

\bibitem{Zu}
 L. Paz and
H. Zurek,
Phys. Rev. Lett. {\bf 82}, 5181 (1999);
 H. Zurek, Annalen der Phys. ???? (2000),
and Lecture delivered at the
conference {\em Hundred Years Quantum Theory\/} in Berlin, 2000.

\end{thebibliography}
\end{document}